\documentstyle[12pt]{article}

\newcommand{\beq}{\begin{equation}}
\newcommand{\eeq}{\end{equation}}
\newcommand{\al}{\alpha}
\newcommand{\be}{\beta}

\newcommand{\di}{\displaystyle}
\newcommand{\vp}{\varphi}
\newcommand{\p}{{\bf p}}
\newcommand{\x}{{\bf x}}
\newcommand{\bl}{\bigl\langle}
\newcommand{\br}{\bigr\rangle}
\newcommand{\J}{\hat J}

\title{Quantum gravitational processes in a hot ultrarelativistic gas 
and their effect on the isotropic Universe evolution.}
\author{Yu. Grishkan \and E. Vertogradova}
\date{Department of Physics, Rostov State University \protect\\
5 Zorge st., Rostov-on-Don, Russia, 344090}

\begin{document}
\maketitle

\begin{abstract}
The variant of quasiclassical (half--quantum) theory of gravity 
in strong gravitational field is presented. The exact solution 
of the problem of the renormalized  energy-momentum tensor calculation is
performed in terms of non-local operator-signed function.
The procedure of quasilocalization is proposed, which leads to 
the equations of non-equilibrium thermodynamics for 
temperature and curvature.
The effects of induced particle creation and media polarization
are taking into account and 
used to solve the problem of non-Einstein's  branches damping.
The problem of Universe  creation from ``nothing'' is also 
discussed.
\end{abstract}

\section{Introduction}
It is well known that in a strong gravitational field $R/m^2\gg 1$ the
effects of 1) vacuum polarization, 2) real particle creation, and 3)
interaction between real particles and self-consistent gravitational field
 take place \cite{1}, \cite{2}, \cite{3}. 
We would like to emphasize that in a gravitating medium no only vacuum 
polarization, but also the polarization of medium occurs, and 
the particle creation may be no only spontaneous, but the induced one.

The purposes of our research are both to conduct the quantitative analysis
of these phenomena and to discuss their influence on the cosmological  
evolution of space-time geometry and of hot ultrarelativistic medium.
We  show, that, being considered together, the polarization of medium and 
the induced particle creation solve the problem of non-Einstein's branches 
damping. Such branches, as it is shown in \cite{4}, \cite{5}, are the 
solutions of Einstein's equations with radiative corrections including
the higher derivatives of metric.

\section{Method of calculation of quantum gravitational corrections
to the Einstein's equations in a strong gravitational field.}

The subjects of our calculations are the Green functions (GF)
of quantum fields in the coinciding space-time points.
For example, in the case of scalar field $\vp(\x, t)$,
this  function renormalized by the Pauli-Willars method is defined as
\beq
D_{ren}(\x, t)=\bl \vp(\x, t) \vp(\x, t) \br_{ren} =
                 \bl \vp(\x, t) \vp(\x, t) \br_{T}-
\sum_a c_a \bl 0| \Phi_a(\x, t) \Phi_a(\x, t)|0 \br.
\label{1} \eeq
Here $a$ run over a number  of  Pauli-Willars's fields,
$c_a$ are the renormalized coefficients;
$\bl 0|$ is Heisenberg's vacuum state vector of auxiliary 
Pauli-Willars's fields; $\bl \dots \br_T$ is the density matrix averaging 
in Heisenberg's representation. Both vector $\bl 0|$ and matrix 
$\bl \dots \br_T$  are defined in some initial time moment. 

In the isotropic Universe
\beq
ds^2=a^2(\eta)\left(d\eta^2-dl^2\right) 
\label{2}
\eeq
one can use the Fourier image of Heisenberg's quantum fields 
$\vp_{\p}(\eta)$ and to introduce the Fourier image of GF:
\beq  
D_{\p}(\eta)=\bl \vp_{\p}^+(\eta)\vp_{\p}(\eta)\br_T.
\label{3} 
\eeq 
Then GF (\ref{1}) is  transformed to 
\beq 
D_{ren}(\eta)=\sum_{\p}D_{\p}(\eta)-\sum_a c_a \sum_{\p} D_{\p a}(\eta) 
\label{3a} 
\eeq
and the task of radiative correction calculation is divided 
into several steps:

(i) to obtain the solution for the Fourier images $D_{\p}(\eta)$;

(ii) to integrate  $D_{\p}$ over momentums according to (\ref{3a});
this gives the corrections to GF.

(iii) to calculate the energy-momentum tensor (EMT) as a functional of 
the renormalized GF.  
 
We have  calculate the exact solutions for GF and EMT in the form of 
the operator-signed function, which depend from operator $\J$:
\[ \J f=\frac14\int\limits_{\tau_0}^{\tau}\frac1{r^2(\tau_1)}
\frac{d^3}{d^3\tau_1}\left(\frac1{r^2(\tau_1)}
           f\right)  d\tau_1, \]
where $f$ is an arbitrary function, time $\tau$ is defined as
$r^3 d\tau=r d\eta=dt$. 
The analytic functions of operator $\J$ are defined as the Teylor series;
but the calculation shows that both GF and EMT are non-analytic 
functions. This fact necessarily shows the presence of non-local effect,
which will be discussed below. 

The formally exact expressions for GF and EMT are

\beq \begin{array}{r} \di 
\pi^2 D_{ren}(\tau)=\frac1{8r^2}\J\ln\J
                 -\frac{5}{24} \frac{r' {}^2}{r^8}+
       \frac{5}{48}\frac{r''}{r^7}-\frac18\ln(\mu^2 r^2)
     \left(\frac{r'^2}{r^8}-\frac12\frac{r''}{r^7}\right)
\\[5mm] \di
+\frac1{8\pi^2}\int\limits_0^{\infty}
                \frac{p^3}{p^2+\J} \frac{dp}{e^{p/\theta}-1}
\end{array}
\label{4}
\eeq

\beq \begin{array}{l} \di
\pi^2 \bl T_0^0\br_{ren}(\tau)=
             \frac1{32 r^6}\left(\frac1{r^2}\J\ln\J\right)''
         -\frac1{8r^4}\J^2\ln\J
+\biggl(-\frac38\frac{r'^4}{r^{16}}
       +\frac14\frac{r'^2r''}{r^{15}}+\frac1{64}\frac{r''^2}{r^{14}}
\\[5mm]
\hspace*{55mm} \di
-\frac1{32}\frac{r'r'''}{r^{15}}\biggr)\ln(\mu^2 r^2)
+\frac{121}{960}\frac{r'4}{r^{16}}
\\[5mm]
\hspace*{15mm} \di
+\frac1{32\pi^2 r^6}\left[ r^4 \int\limits_0^{\infty}
           \frac{p^3}{p^2+\J} \frac{dp}{e^{p/\theta}-1}\right]''
+\frac1{8\pi^2r^2}\int\limits_0^{\infty}
        \frac{p^5}{p^2+\J} \frac{dp}{e^{p/\theta}-1}.
\end{array}\label{5}
\eeq
Here  prime  denotes $'={d}/{d\tau}$;  
      $\theta=Ta$ is a conformal temperature of a medium; 
      $\mu$ is renormalizaion scale.

The terms in (\ref{4}), (\ref{5}), which do not contain the 
Bose-Einstein function, describe the vacuum effects; the terms, which 
contain this function, describe the quantum gravitational effects
in the real particle sector. 
Two types of nonanalyticities  are present in equations (\ref{4}),
(\ref{5}): those in vacuum sector are logarithms from operator $\J$;
those in real particle sector are integrals containing operator
$\J$ and Bose--Einstein's distribution function.
We consider vacuum and "matter" terms together and made the asymptotic
expansion of the integrals on parameter $\di \left({\J}/{T^2}\right)^{1/2}$,
then the nonanalyticities of $\ln\J$ type are canceled.
All terms  $F_n=\di \left({\J}/{T^2}\right)^{n/2}$, $n=0,1,2,3,\dots$ 
in asymptotic expansion
 are  divided into two type. For even $n$
all $F_n$ are local functions of metric, for odd $n$ they are
principally non-local.
For example, the linear on temperature term in EMT is
\beq
\bl T_0^0\br_{\theta}=\theta\left[
-\frac{1}{16\pi^2r^2}\left(\frac1{r^2}\J^{1/2}\right)''+
     \frac1{4r^4}\J^{3/2}\right]
\label{6}
\eeq
This term and other odd on temperature terms, being non-local ones, 
describe the effect of the induced creation of real particles.
In the next sections we work in the framework of nonequilibrium 
thermodynamics and so we develop a special method to make a 
quasilocalization of term (\ref{6}) and similar ones.

\section{The problem of quasilocalization of nonanalytic 
operator-signed functions.}

Notice, the above described calculations are made when conformal temperature
$\theta$ is supposed to be constant. The procedure of quasilocalization
means the nonlocal dependence would be transferred from 
operator $J$ degrees to the conformal temperature, which became
irreversibly dependent from time $\tau$. At the first step 
we find  approximations
\beq 
\begin{array}{l} \di
\frac{1}{4\pi^2}\left(\frac{1}{r^2}\J^{1/2}\right)''\approx
\left(\frac1{r^2}\right)' \left(\frac1{r^2}\right)''
+\frac18\left(\frac1{r^2}\right) \left(\frac1{r^2}\right)'''
-\frac1{27}\left(\frac1{r^2}\right)'{}^3\cdot r^2 
\\[5mm] \di
\frac{4}{\pi^2}\left(\frac{1}{r^2}\J^{3/2}\right)\approx
\frac13\left(\frac1{r^2}\right)'{}^3
+\frac12\left(\frac1{r^2}\right)'''
+\frac12\left(\frac1{r^2}\right) \left(\frac1{r^2}\right)' 
                                \left(\frac1{r^2}\right)''
+\frac1{18}\int\limits_{\tau_0}^{\tau} \sqrt{-g} R^2 d\tau
\end{array}
\label{7}
\eeq

\beq
\bl T_0^0\br_{\theta}\approx\frac{\pi^3}{32}\theta\left[
-\left(\frac1{r^2}\right)^3
         \left(\frac1{r^2}\right)' \left(\frac1{r^2}\right)''
+\frac19\left(\frac1{r^2}\right)^2
\int\limits_{\tau_0}^{\tau} \sqrt{-g} R^2 d\tau \right]
\label{8}
\eeq
In equations (\ref{7}),  (\ref{8}) $\theta=\theta(\eta_0)=const$ is 
the initial conformal temperature, but the integral term  describe
the particle creation.
At the second step the non-local time dependence transforms to 
the conformal temperature irreversible time dependence 
$\theta=\theta(\eta)$. In order to vanish integral from EMT the law 
of entropy increase should be as follows:
\beq
\dot\theta \theta^2=\frac{1}{\pi^2}\frac{k_3}{k_1}\frac{\ddot r^2}{r^2},
\label{9}
\eeq
where the dot denotes $\frac{d}{d\eta}$ , the coefficients $k_1$ , $ k_3 $
see below.

Notice, the proposed way of localization  of operator-signed functions
is the simplest one, but other ways are not excluded also. This question
is the discussible one.

As a result Einstein's equation with the renormalized EMT  
\[
R_0^0-\frac12R ={\rm\ae}\bl T_0^0 \br_{ren}
\]
in the isotropic space--time with metric (\ref{2}) 
has a form
\beq 
\frac1{l_{Pl}^2}\dot r^2=
       \beta\frac{\dot r^4}{r^4}+k_1\theta^4+k_2\frac{\dot r^2}{r^2}
+k_3\theta\left(\frac{\dot r^3}{r^3}-\frac{\dot r\ddot r}{r^2}\right)
-k_4\left(2\frac{\dot r \,  r\hspace*{-2mm}{}^{^{\textstyle ...}} }{r^2}-\frac{\ddot r^2}{r^2}
          -4\frac{\dot r^2 \ddot r}{r^3} \right) 
\ln\frac{\mu^2 r^2}{\theta^2}.
\label{11} \eeq
For one scalar field the coefficients are
\[ \be=\frac1{1440\pi^2}, \quad
k_1=\frac{\pi^4}{30}, \quad k_2=\frac{\pi^2}{24}, \quad k_3=\frac{\pi^3}{32},
\quad k_4=\frac{9}{8\pi^4}. \]

Equations (\ref{9}), (\ref{11}) form the closed self-consistent
system for the cosmological variables $r(\eta)$ and $\theta(\eta)$.

\section{Damping of non-Einstein branches of cosmological solutions.}

The system (\ref{9}), (\ref{11}) of differential equations is solved,
utilizing the theory of continuous symmetry groups \cite{4}, \cite{5}. 
At the first step one should reduce the order of this system; at the 
second step the WKB-solution on small parameter --- dimensionless
Hubble constant 
\beq 
h=H l_{Pl}=\frac{\dot r}{r^2}l_{Pl}\ll1
\label{12}
\eeq
is found:
\beq \begin{array}{rr} \di
\ln\frac{r}{r_0}=-\frac12\ln h- A\left(\frac{k_4}{h}\phi\right)^{1/4}
\exp\left(-\frac14\frac{k_3}{k_4 \sqrt{k_1}}\phi\right)\cdot
\cos(\phi-\phi_0); 
\\[4mm] \di
\frac1{l_{Pl}}(t-t_0)=\frac12 h- A h \left(\frac{k_4}{h}\phi\right)^{1/4}
\exp\left(-\frac14\frac{k_3}{k_4 \sqrt{k_1}}\phi\right)\cdot
\cos(\phi-\phi_0); 
\end{array} \label{13}
\eeq
here 
\[\phi=\di\frac1{4h}\cdot\frac1{\di\ln\frac1{h}}>0, 
                  \qquad A,\;\phi_0 \mbox{ --- integration constants} \]
It is easy to see that the first terms in (\ref{13}) represent
the Friedmann-Robertson-Walker (FRW) solution for radiative-dominant 
plasma $r\sim t^{1/2}$. The second terms correspond to 
the non-Einstein branches. The problem is: 
for the existence of the observed FRW Universe  non-Einstein branches 
must be damped. From our general solution (\ref{13}) it is obvious that
the key role is played by the parameters $k_3$, $k_4$. 
Being introduced in EMT (right side of Einstein's equations (\ref{11})), 
they describe the 
effects of induced particle creation and
matter polarization in the Universe.

Now it is clear why the problem of non-Einstein's  branches was 
unsolved until now.  Usually  in quasiclassical theory the key 
effects --- induced particle creation and
matter polarization were not taken into account (see for example 
\cite{6}, \cite{7}). (Formally this case correspond to $k_3=0$.)
In this case in (\ref{13}) $h\to 0$ as $t\to \infty$,
the exponent is absent, the pre-exponential multiplier 
is leading term and non-Einstein's
branches grow up as $\sim 1/h^{1/2}$; 
their oscillations  predominate over FRW terms
--- this result was obtained numerically (see, for example, \cite{6}).
We show it is of great value that $k_3>0$, and non-Einstein's  branches
are exponentially damped $\sim 1/h^{1/2}\exp(-k_3/h)$.

Our consideration show 
that the problem of non-Einstein branching damping do not need to use
any exotic ideas like n-dimension gravity.
The problem can be solved by  punctual account of all matter 
effects (particularly, both spontaneous and induced particle creation;
and vacuum and  medium polarization)  in the 
framework if usual quasiclassical (half-quantum) theory of gravity.

\section{The problem of Universe creation from ``nothing''.}

We assume that the obtained equations (\ref{9}), (\ref{11})
can be used for the formal description of early stages of Universe 
evolution. Particularly, this description performs to solve
the problem of Universe creation from ``nothing'', here ``nothing''
means the physical state with $R=0$, $T=0$.  

We found the assimptotical at $R\ll R_{Pl}$, $T\ll T_{Pl}$ solution
of equations (\ref{9}), (\ref{11}) in the form:
\beq
\frac{r}{r_0}\sim \exp\left[\al \frac{(t-t_0)^3}{l_{Pl}^3}\right],
\label{14}
\eeq
here $r_0$ , $\al$ , $t_0$ are  arbitrary constants. 
This equation corresponds to the laws of curvature and temperature
increasing 
\par
\beq 
\begin{array}{l}
R_0^0-\frac12R \sim (t-t_0)^4,   \\[4mm]
T \sim t-t_0,
\end{array}
\eeq
such increasing means the particle creation in early Universe.
It's obvious that curvature $ R^0_0 $ and temperature $ T $ are
equal to zero at the initial time moment:
\[      
(R_0^0-\frac12R)|_{t_0} = 0, \qquad T|_{t_0}=0
\]
This solution at $t=t_0$ corresponds to the empty Minkowski space.
In quasiclassical theory just this state can be interpreted as
``nothing''. 
During the evolution this initial state decays to geometry 
and particles under the action of "initial push", which in mathematical
terms is described as nonzero fourth derivative: 
\par
\begin{equation}
\left(R^0_0-\frac{1}{2}R\right)^{(4)} |_{t_0} \ne 0. 
\label{15} 
\end{equation}
The assumption about thermodynamic equilibrium at early stages of
cosmological evolution used in (\ref{9}), (\ref{11}), physically 
corresponds
to the hypothesis about quick termalization of the created 
particles. This hypothesis is obviously rough because in the vicinity 
of a singularity near the Planck space-time scales the characteristic time
of particle creation and the relaxation time of their gas are 
comparable. From this point of view the model is phenomenological,
but nevertheless it involves all 
the main points of the discussed phenomena.

To clarify the whole phase picture of model (\ref{9}), (\ref{11}), 
and to count a whole number of the created particles $ N |_{t=\infty} $ it's
necessary to found a whole set of analytical asymptotics  and to
make the numerical integration of system (\ref{9}), (\ref{11}). 
This work is in progress.

\vspace*{0.7cm}

We thank G.M. Vereshkov, O.D. Lalakuulich for helpful discussions
and critical comments.

\end{document}